\documentclass[twocolumn, switch]{article} 

\usepackage{preprint}

\usepackage{amsmath, amsthm, amssymb, amsfonts}
\usepackage{dirtytalk}
\usepackage{algorithm}
\usepackage{algpseudocode}

\usepackage[numbers,square]{natbib}
\usepackage[utf8]{inputenc}	
\usepackage[T1]{fontenc}	
\usepackage{xcolor}		
\usepackage[colorlinks = true,
            linkcolor = purple,
            urlcolor  = blue,
            citecolor = cyan,
            anchorcolor = black]{hyperref}	
\usepackage{booktabs} 		
\usepackage{nicefrac}		
\usepackage{microtype}		
\usepackage{lineno}		
\usepackage{float}			

\usepackage{lipsum}		

\usepackage{newfloat}
\DeclareFloatingEnvironment[name={Supplementary Figure}]{suppfigure}
\usepackage{sidecap}
\sidecaptionvpos{figure}{c}

\usepackage{titlesec}
\titlespacing\section{0pt}{12pt plus 3pt minus 3pt}{1pt plus 1pt minus 1pt}
\titlespacing\subsection{0pt}{10pt plus 3pt minus 3pt}{1pt plus 1pt minus 1pt}
\titlespacing\subsubsection{0pt}{8pt plus 3pt minus 3pt}{1pt plus 1pt minus 1pt}

\usepackage{tikz,xcolor,hyperref}

\definecolor{lime}{HTML}{A6CE39}
\DeclareRobustCommand{\orcidicon}{
	\begin{tikzpicture}
	\draw[lime, fill=lime] (0,0) 
	circle [radius=0.16] 
	node[white] {{\fontfamily{qag}\selectfont \tiny ID}};
	\draw[white, fill=white] (-0.0625,0.095) 
	circle [radius=0.007];
	\end{tikzpicture}
	\hspace{-2mm}
}
\foreach \x in {A, ..., Z}{\expandafter\xdef\csname orcid\x\endcsname{\noexpand\href{https://orcid.org/\csname orcidauthor\x\endcsname}
			{\noexpand\orcidicon}}
}

\title{Content-aware media retargeting based on deep importance map}

\usepackage{authblk}

\author[1]{Thi-Ngoc-Hanh Le}
\author[2]{Shih-Syun Lin}
\author[3]{Weiming Dong}
\author[1]{Tong-Yee Lee}

\affil[1]{National Cheng-Kung University, Taiwan}
\affil[2]{National Taiwan Ocean University, Taiwan}
\affil[3]{Chinese Academy of Science and School of Artificial Intelligence, China.}

\begin{document}

\twocolumn[ 
  \begin{@twocolumnfalse} 
  
\maketitle

\begin{abstract}
We present a neural network to estimate the visual information of important pixels in image and video, which
is used in content-aware media retargeting applications. Existing techniques are successful in proposing retargeting methods. Yet, the serious distortion and the shrunk problem in the retargeted results still need to be investigated due to the limitations in the methods used to analyze visual attention. To accomplish
this, we propose a network to define the importance map, which is sufficient to describe the energy of the
significant regions in image/video. With this strategy, more ideal results are obtained from our system. Besides,
the objective evaluation presented in this paper shows that our media retargeting system can achieve better
and more plausible results than those of other works. In addition, our proposed importance map
performs well in the enlarge operator and on the \say{difficult-to-resize} images.
\end{abstract}
\vspace{0.35cm}

  \end{@twocolumnfalse} 
] 

\section{Introduction}
Media retargeting has been an active and attractive research topic in both computer vision and computer graphics in recent years due to the evolution of heterogeneous devices for capturing images and videos as well as for displaying them. Their displays also have different resolutions and aspect ratios. Thus, methods for changing the sizes of images and videos are gaining importance. A content-aware image/video retargeting method has been mentioned as an algorithm which focuses on the aspect that important objects in the image should be preserved when it is retargeted to other display resolutions.

\begin{figure}[hbt!]
  \centering
  \includegraphics[width=3.0in]{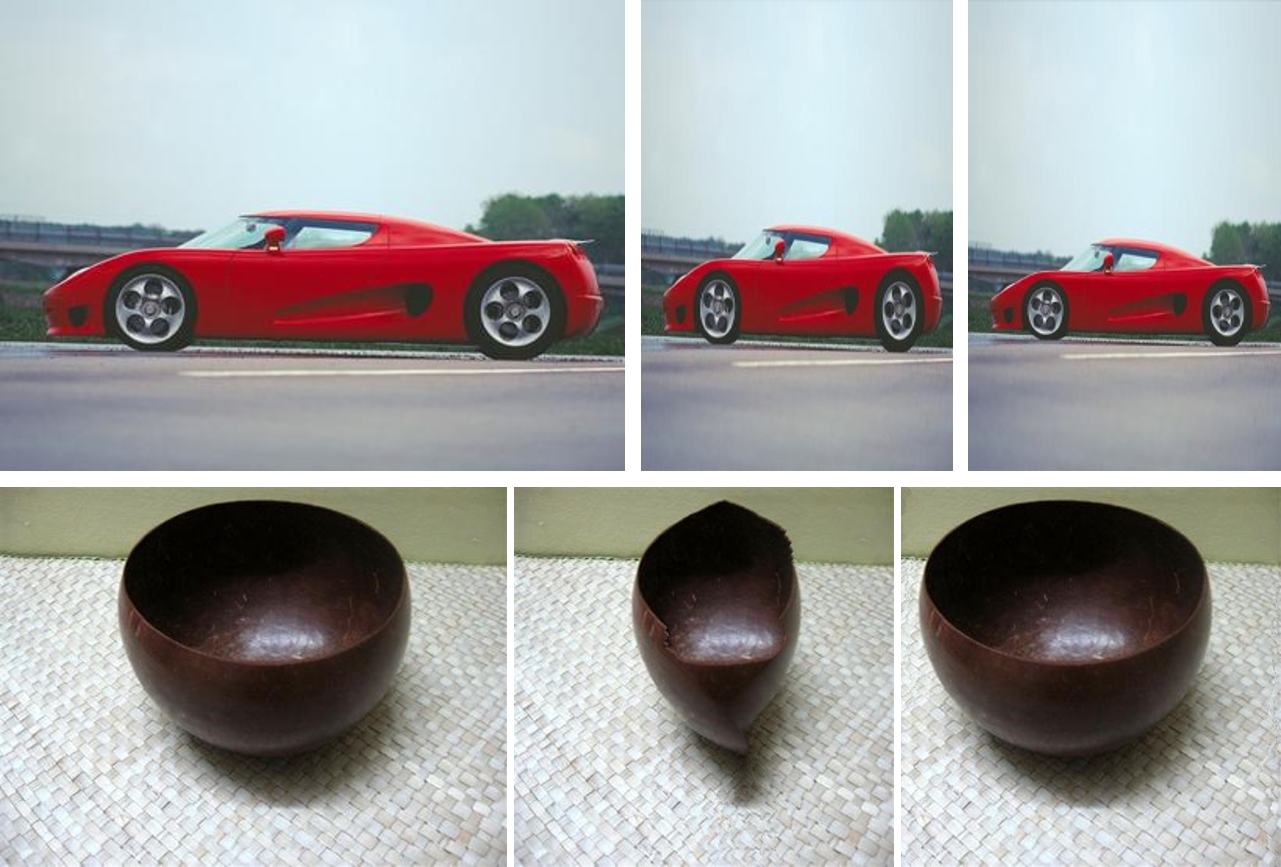}
  \caption{Visualization of the limitations in the previous results. The first row, from left to right: original image, \citet{lin2012patch}'s result, our result. The second row, from left to right: original image, \citet{avidan2007seam}'s result, our result.}
  \label{f_example}
\end{figure}

The conventional content-aware image/video retargeting methods \cite{avidan2007seam, jin2010nonhomogeneous, rubinstein2009multi,pritch2009shift, lin2012patch, lin2013content} typically rely on the visual information of image/video to define the importance in image/video which should be preserved after retargeting. The objective of a retargeting method is to preserve either the importance of image or the ratio of objects. These approaches have been successful in maintaining the importance of image/video and hindering the distortion in the retargeted output. However, they solely focus on designing a good retargeting technique. The methods, which they use to identify importance in image/video, are not sufficient to describe the visual information of the significant regions. Such as the results in \cite{avidan2007seam} are distorted seriously with the gradient map. Or, \citet{lin2012patch, lin2013content} generate shrunk results due to the lack of energy on the important regions (as the examples shown in Fig.\ref{f_example}). Some recent works \cite{cho2017weakly, song2018carvingnet, tan2019cycle} attempt to adopt deep learning techniques to work on the problems in this research field. However, the retargeting results are still need to investigated. Besides, the challenge in building a large image retargeting dataset asks for the retargeted images not only to maintain what people look forward to but also to achieve aesthetic quality demands.

Our current work is motivated by seam carving and warping based method which is studied in \cite{avidan2007seam, lin2012patch, lin2013content}. We propose a network to estimate the importance of image/video frames. It is then employed in a comprehensive framework to produce retargeting results. With our system, we tackle the aforementioned drawbacks and boost the retargeting results more ideal. Our contributions can be summarized in the following issues:
\begin{itemize}
\item [$\bullet$]We introduce a comprehensive system which is able to produce the better media retargeting results for seam carving-based and warping-based method.
\item [$\bullet$]We adopt a deep learning method to estimate the important pixels of a given image/video which is used for image and video retargeting applications. 
\item [$\bullet$]More plausible retargeted images and videos are obtained from our proposed system.
\end{itemize}

We organize the remainder of this paper as follows. Section 2 is a brief review on the related studies. Section 3 presents the proposed network and explains how our importance map tackles the problems in image and video retargeting. In Section 4, the evaluation on the results is described and a discussion on our results follows. Finally, section 5 presents the conclusion and our future work.

\section{Related work}
In this section, previous studies, which are close to our current approach, are reviewed. They consist of (1) the media retargeting techniques including conventional approaches and deep learning-based approaches, and (2) the schemes that are proposed to improve media retargeting results.

\begin{figure}
  \begin{center}
  \includegraphics[width=3.0in]{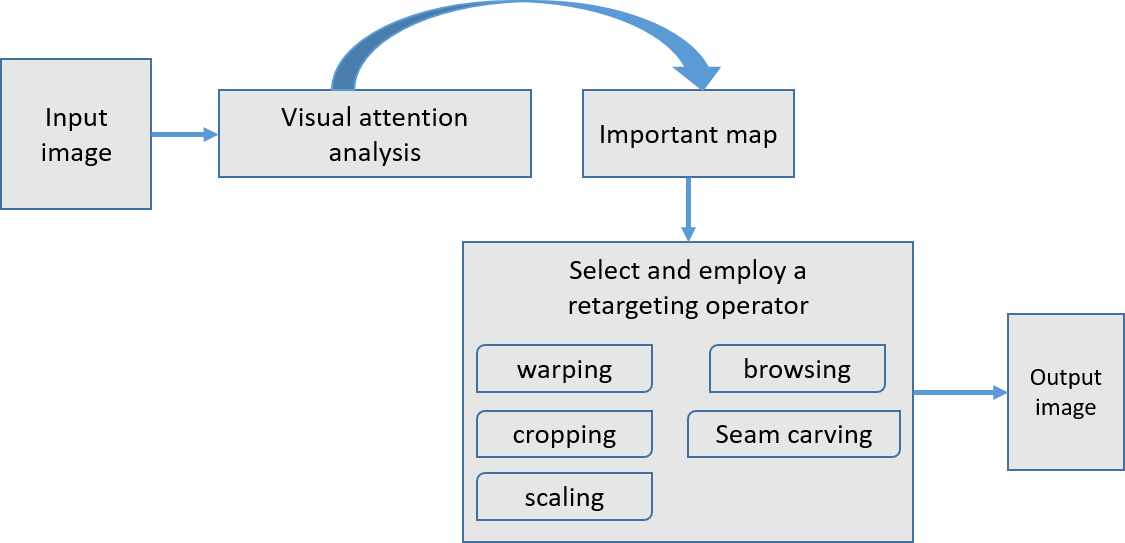}\\
  \caption{The basic workflow of the content-aware media retargeting frameworks \cite{kiess2018survey}.}
  \label{basic_workflow}
  \end{center}
\end{figure}

The existing approaches for Content Aware Image Retargeting (CAIR) are probably categorized into discrete and continuous methods \cite{kiess2018survey}. In both two categories, the pipeline for retargeting firstly extracts the importance from the input image/video frames then utilizes these information to drive the input image/video frames to the target size by an resizing operator (as demonstrated in Fig.\ref{basic_workflow}). Relying on this workflow, importance in the input image is considered as a required process in a resizing system proposed by \cite{suh2003automatic, kopf2009adaptation, avidan2007seam, rubinstein2009multi, pritch2009shift, lin2012patch, lin2013content}. These typical resizing methods utilize different methods to determine the image importance in their own framework.

A naive approach was early used in resizing an image in terms of cropping. The main challenge of cropping is finding an ideal cropping window. To overcome this challenge, cropping techniques begin by calculating an importance map. This map is obtained from a saliency method that is introduced in \cite{itti1998model}.
Seam carving \cite{avidan2007seam} has been mentioned as the first method in content-aware image retargeting. The main idea of this method is to remove seams that consist of low-importance pixels to drive an image to its target size. To estimate the weights of pixels, this approach utilizes the gradient-based energy map of the input image and removes a seam where the energy is minimum. Inspired by this concept, the studies in \cite{rubinstein2008improved, han2010optimal, ito2016gradient, yin2015detecting} are lately introduced to improve the fundamental concept. Most of these works perform well in landscape images or pictures with large homogeneous areas. However, in the cases that images consist of straight line, regular shapes or numerous objects, visual artifacts can occur in \cite{kiess2018survey}. Besides, all less-important regions have been removed leads to serve artifacts.

While scaling resizes the entire image uniformly by the same scale factor, warping is content-adaptive and determines scaling factors in a fine-grained way \cite{kiess2018survey}. Proposed in \cite{liu2005automatic}, this study tries to preserve important content in an image. To measure the importance, an importance map is created consisting of a saliency map computed by \citet{ma2003contrast}. Yet, image features outside the region of interest may suffer from significant distortions.

For the warping-based approach, \citet{zhang2009shape} proposed a method with the attempts to ensure that important local regions undergo a geometric similarity transformation as well as preserve image edge structure. To achieve this, an importance map of the source image is first defined. Based on it, each important region is assigned a value. \citet{guo2009image} presented an image retargeting method using a mesh image representation. A saliency map of the source image is first produced by method proposed in \citet{ma2003contrast}. It is then associated with the source mesh. In this way, retargeting is transformed into a parametrization problem of finding a target mesh with the desired resolution. \citet{jin2010nonhomogeneous} introduced an approach for interactive content-aware image resizing. The image saliency information of image is captured to define visual information in the input image. A triangle mesh is then created. The scales of all triangles are obtained by solving an optimization problem. To preserve visually salient objects and structure lines in retargeting an image, \citet{lin2012patch} proposed a novel scheme in term \say{patch-based}. In this framework \citet{lin2012patch}, input image is first segmented to partion image. A saliency map method \cite{goferman2011context} is then employed. Each segmented patch is assigned an average saliency value. However, this method can not work well for the images contain with similar significant values \cite{lin2012patch}.

In term of video retargeting, \citet{wolf2007non} early introduced an efficient algorithm for video retargeting. To detect the importance of each region in frames, frames extracted from the input video are first analyzed. The proposed analysis is fully automatic and based on local saliency, motion detection and object detectors. A transformation is then applied with the respects that the analysis shrinks less important regions more than importance ones. Motivated by seam carving method, \citet{rubinstein2008improved} proposed to improve seam carving method in video retargeting. Instead of removing 1D seams from 2D images, they remove 2D seam manifolds from 3D space-time volumes. To reduce the unpleasant distortion, \citet{lin2013content} presented a novel continuous approach in video retargeting. This approach proposed an object-preserving warping scheme with object-based significance estimation. In this scheme, visual salient objects in 3D space-time space are forced to undergo as-rigid-as-possible warping, while low significance content are warped as close as possible to linear scaling.

From above, even each system utilizes a different method to define the importance in image/video frames, this process is considered as a prerequisite stage in such retargeting framework. With the attempt to boost the retargeting results better, the following methods improved the drawbacks in previous works by different ways. To avoid artifacts in the conventional seam carving, \citet{achanta2009saliency} introduced a saliency dectection method. Unlike gradient maps, which may have to be recomputed several times during a seam carving based retargeting operation, saliency maps in \citet{achanta2009saliency} are computed once independent of the number of seams added or removed. \citet{kim2011spatiotemporal} developed a visual saliency measure to identify the important image regions for obtaining the image importance map. They propose a robust importance measure of self-ordinal resemblance (SOR) defined by computing distances between ordinal signature of edge and color orientation histograms obtained from center and surrounding regions. After calculating image importance map, they employ the improved seam carving \cite{rubinstein2008improved} for retargeting. \citet{wang2014salient} presented salient edge and region aware image retargeting (SERAR) method, by decomposing an image into a cartoon and a texture part to provide the reliable salient edges from cartoon part. Based on the proposed image importance map, they modified the seam carving to avoid distorting the geometric structures.

Recently, with the advances in learning-based approaches, researchers have attempted to use convolutional neural network to define the importance in images. \citet{song2018carvingnet} proposed an improved content-aware image resizing method that uses deep learning. The proposed method is extended from seam carving. Instead of using gradient based to compute the energy map, they propose a method for creating a deep energy map using an encoder-decoder convolution neural network. To improve retargeting results in both saliency detection and retargeting, \citet{ahmadi2019context} presented a new approach in which the use of image context and semantic segmentation are examined.

In contrast, we propose a neural network scheme to define the importance map in both image and video frames. This map is then examined by any retargeting operator. Our experimental results and the evaluation presented in this paper show our approach is effective in improving media retargeting results.
\begin{figure*}
  \centering
  \includegraphics[width=1.85\columnwidth]{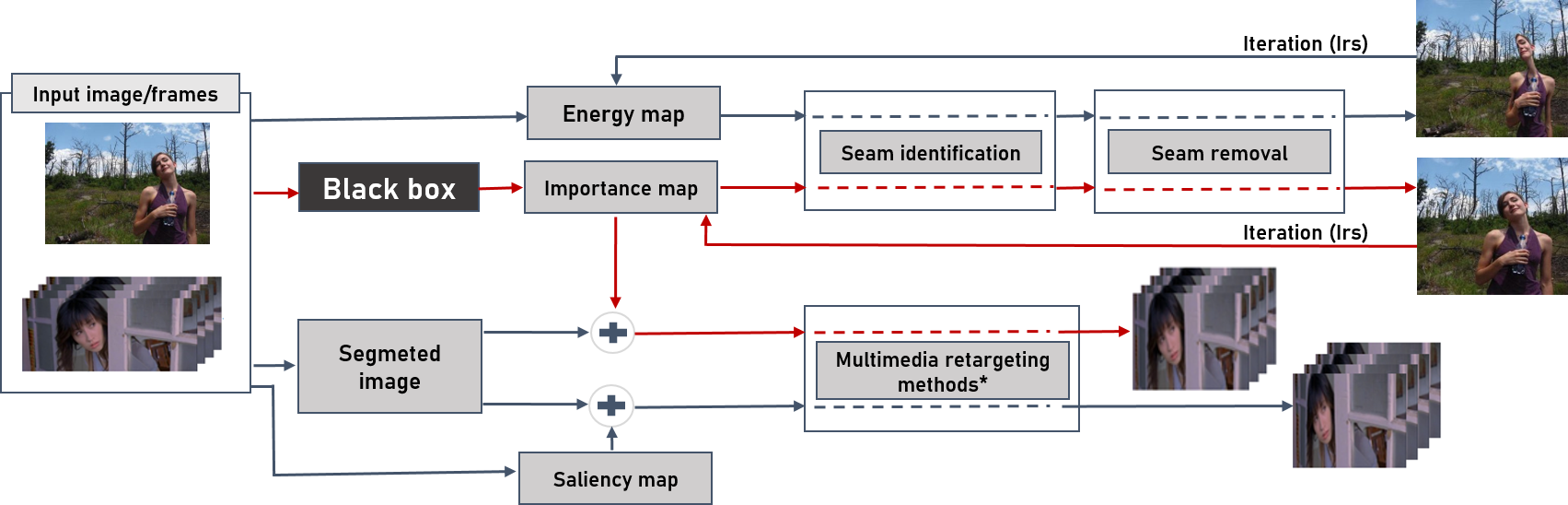}
  \caption{The pipeline of our proposed retargeting system. The black arrow denotes the navigation procedure of the previous system. The red arrow is the navigation procedure of our media retargeting system.}
  \label{f_main}
\end{figure*}

\section{Methodology}
The overall pipeline of our proposed media retargeting system is presented in Fig.\ref{f_main}. Given an image or video, output of our system is an image/video which is retargeted to a new display resolution size. The proposed system consists of two main stages. The first one, which is mentioned as a \textit{\say{black box}}, is to calculate the visual information in image/video through the proposed network. Meanwhile, the remain stage is to produce retargeted image/video by the retargeting operations. In the following, the specification of each process is described.

\subsection{Visual information estimation via the proposed network}
The first fundamental step of a content-aware retargeting pipeline is the estimation of importance of pixels in images/video frames \cite{kiess2018survey}. Therefore, analyzing the visual information to define the important pixels in images/videos is a prerequisite in our proposed system. As mentioned in related work section, the patch-based approach \cite{lin2012patch, lin2013content} utilizes a context-aware saliency method \cite{wolf2007non} to specify either nonhomogeneous regions or homogeneous regions, and seam carving technique \cite{avidan2007seam} adopts the gradient-based calculation to define the removed/inserted seam. Yet, due to the diversity of media in the real world, the aforementioned techniques, which are utilized to produce the importance of pixels, do not perform well. This phenomenon results in retargeted media are not plausible. Thus, preserving the shape of important objects in the retargeted media is a necessary manner to be investigated. Motivated by this, we propose a network to estimate pixel information with the attempt to boost the retargeting media better. The proposed network consists of two phases in the terms of \say{\textit{feature extraction}} and \textit{\say{layer-fusion}}. Feature extraction is built upon a pre-trained network to extract features in an image/video frames. Later, a computation in terms of \say{layer-fusion} is adopted to estimate the important pixels in the given input.

\begin{figure}[hbt!]
  \centering
  \includegraphics[width=3.3in]{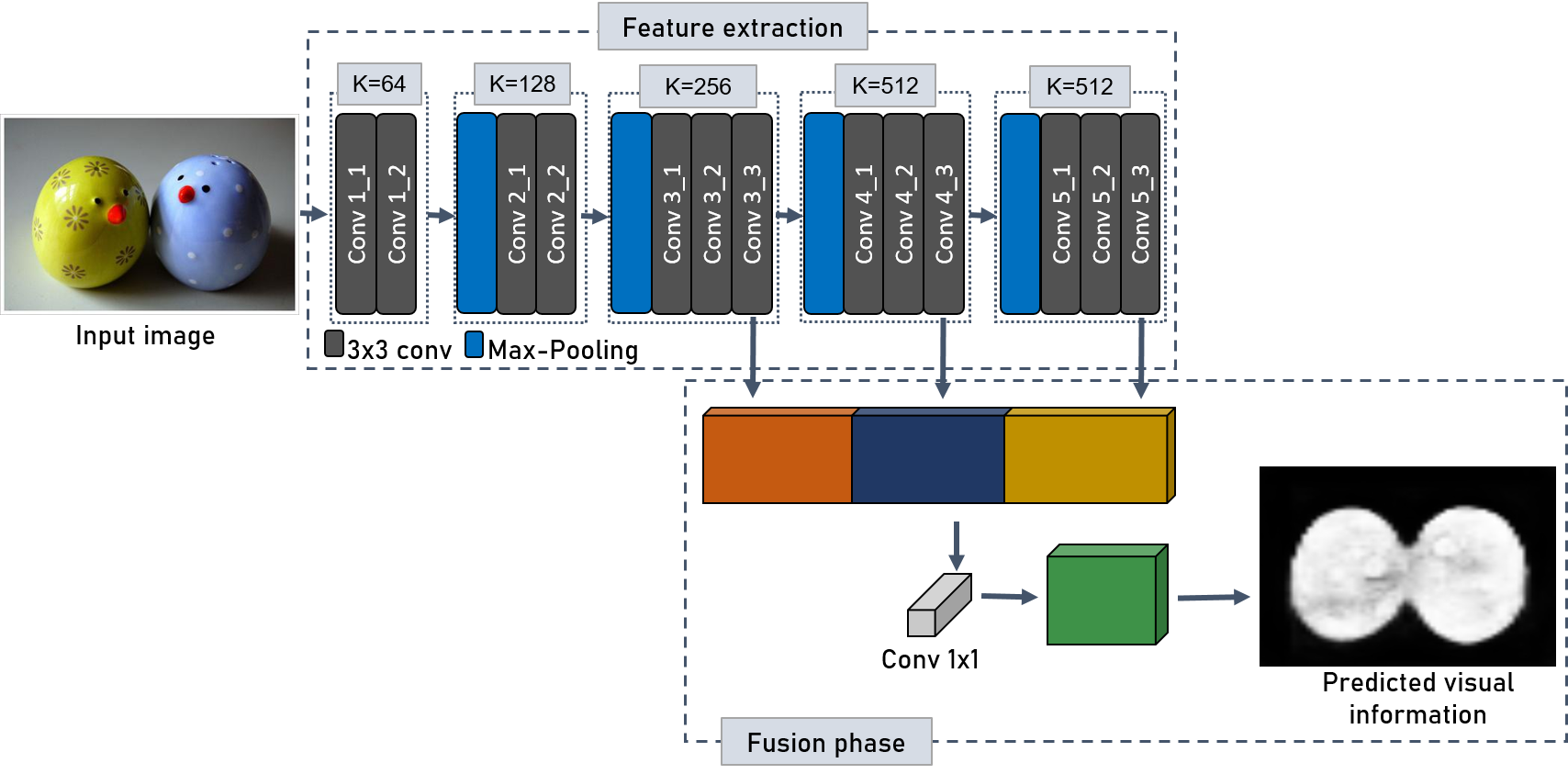}
  \caption{The proposed network.}
  \label{f_framework}
\end{figure}

\textbf{Feature extraction}\newline
To extract features in images/video frames, the pre-trained VGG-16 network \cite{simonyan2014very} is adopted. This well-known network is trained with ImageNet dataset \cite{ILSVRC15} and has excellent performance in image classification, object detection, etc. Besides, there are many of deep learning-based studies utilized VGG-16 network to extract features in their applications. Thus, this pre-trained network is suitable to adequately extract feature for our feature extraction manner. VGG-16 network consists of 13 convolution layers, 5 max-pooling layers, and 3 fully-connected layers follow a stack of convolution layers, and a soft-max layer goes at the end \cite{simonyan2014very}. The fully-connected and soft-max layers are typically used in image classification and object detection applications. In our network, since we take advantage of VGG to extract image/video frame features, we remove either fully-connected or soft-max layers and build our network upon the convolution and pooling layers. These layers form five blocks as we demonstrate in Fig.\ref{f_framework}. Layers in such a VGG network is applied by the different number of filters $(K)$. Feature maps at each layer capture different information from the input. As shown in Fig.\ref{f_framework}, except the first one, each block composes of a max-pooling layer and convolution layers. We assume that the input image size is $W$x$H$x$c$, where $W$, $H$, $c$ is the width, height, and the number of channels, respectively. Due to the utilization of max-pooling layers, feature maps at the lower layers are defined as $(W/2^i$x$H/2^i$x$c), i=0\dots3$.

\textbf{Layer-fusion}\newline
To analyze the visual information of important pixels, we first propose to fuse the feature maps from the layers in our network. Then a tensor obtained from the first manner is used to predict visual information of the input image. Visual attention map for media retargeting application needs to present the contrast color values on both the edge of important objects and the regions inside such these objects. Encoding directly from a single block may not guarantee to obtain the expected feature. Thus, we introduce a strategy in terms \say{layer-fusion}. The fused layers are extracted from the last convolution layer in the $3^{th}, 4^{th},$ and $5^{th}$ block. As the demonstration in Fig.\ref{f_framework}, these layers are named $conv3\_3, conv4\_3,$ and $conv5\_3$, respectively. In the following, we call these maps as, respectively, $F_3, F_4$, and $F_5$. 

\begin{figure}[hbt!]
  \centering
  \includegraphics[width=3.0in]{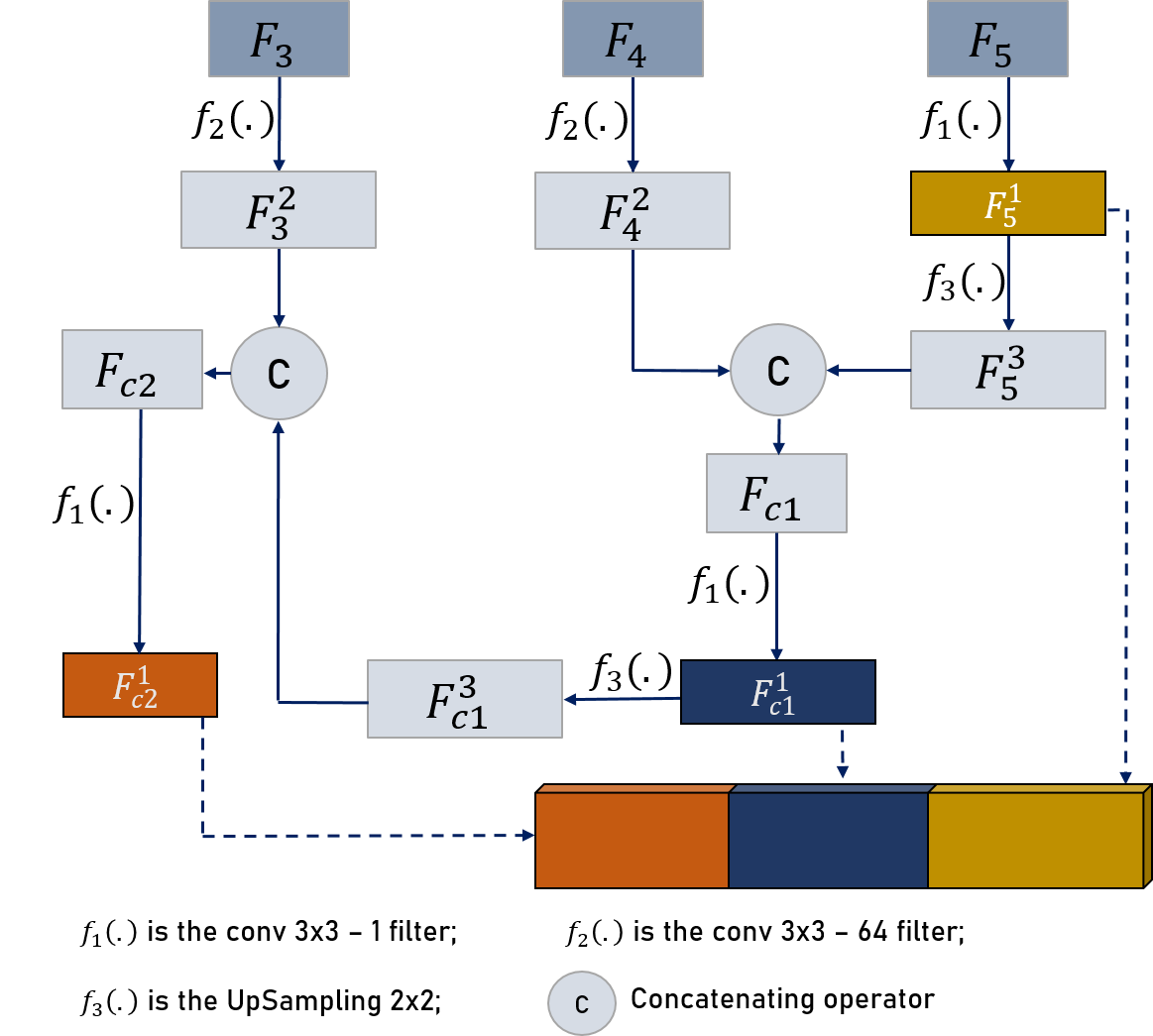}
  \caption{The illustration of the fusion process.}
  \label{f_fusion}
\end{figure}

We solve the fusion through a concatenate operator. Since each pair of feature maps, which are concatenated, are required to share the same spatial size, the following describes our procedure to archive this requirement. Given two adjacent feature maps $F_i$, $F_j$ $(i<j)$, the product of the two $(F_c)$ is defined as:
\begin{equation}
    F_c = f_c(\hat{F_i}, \hat{F_j})
\end{equation}, where $f_c(.)$ is the concatenate operator,
\begin{equation}
    \hat{F_i} = f_2(F_i)
\end{equation}, where $f_2(.)$ is the conv 3x3 - 64 filters, 
\begin{equation}
    \hat{F_j} = f_3(F_j)
\end{equation}, where $f_3(.)$ is the UpSampling 2x2 window. Once the product of a concatenation is obtained, it is fed into a conv 3x3 - 1 filter. Those of such products form a tensor $(F_T)$, which is expressed as:
\begin{equation}
    F_T = f_1(F_{c2}) \otimes f_1(F_{c1}) \otimes f_1(F_5)
\end{equation}, where $f_1(.)$ is the conv 3x3-1 filter, $F_{c2}$ is the product generated by (1) with the input layers are $F_3$ and $F_4$, $F_{c1}$ is the product generated by (1) with the input layers are $F_4$, and  $F_5$. The detail of this procedure is illustrated in Fig.\ref{f_fusion}.  

Finally, a 1x1 convolution \cite{lin2013network} is adopted for mapping the feature maps into a ground truth through an activation function. The ground truth we imply in our network is the masked image corresponding a color image. That is, the important objects in an image are masked in white, whereas the background is in black. In this final layer, we use a sigmoid activation function to calculate the probability as an output that has value in the range of 0 and 1. In the training process, all the parameters in our network are learned by minimizing the loss function, which is computed by the errors between the probability map and ground truth. Given a ground truth $S_g (S_g\in {0,1}^{hxw})$ which is corresponding to the input image I (HxWxc), stochastic gradient descent (SGD) is employed to minimize the loss of training to predict visual information probability: \begin{equation}
    L(S_g, S_p) = -y_i * log(\hat{y_i}) + (1-y_i)*log(1-\hat{y_i})
\end{equation}, where $y_i\in S_g$ and $\hat{y_i}\in S_p$.

The effectiveness of the importance map produced by the proposed network in retargeting results is shown in Fig.\ref{f_result_map}. Information of image is visualized by the corresponding heat map. If the important pixels are defined insufficiently (Fig.\ref{f_result_map}-b), retargeted image is shrunk (Fig.\ref{f_result_map}a-1). Meanwhile, with a correct importance map (Fig.\ref{f_result_map}-c), better preservation of the shape of important objects in the retargeted image is generated (Fig.\ref{f_result_map}a-2). More experimental results in retargeting image and video are shown and discussed in later session.


\subsection{Media retargeting}

Once the importance map is produced, the second step of our retargeting pipeline is employing retargeting operation based on the generated importance map. In this paper, we examine two approaches to produce media retargeting results. One of them represents for the continuous methods. We use the modified version of the study in \cite{lin2012patch} for image resizing and \cite{lin2013content} for video retargeting. The other one, which has been mentioned as a typical technique in discrete category, is seam carving technique in \cite{avidan2007seam}.

\textbf{Image retargeting with seam carving method}\newline 
To drive an image to a target size, \citet{avidan2007seam} relied on an energy map to define a vertical/horizontal seam which is removed/inserted from/into the input image. We describe this technique in reducing image size as the below pseudocode:
\begin{algorithm}
\caption{Algorithm of seam carving}
	\text{Require: An RGB image $I_o, Irs$.}
	\begin{algorithmic}[1]
		\For {$i=0$ to $Irs$}
		    \State $E_i = E(I)$; 
		    \State $s_i$ $\xleftarrow[]{}$ the minimum seam;
		    \State $i^{++}$;
		    \State $I_i$ $\xleftarrow[]{}$ Remove$(E_i, s_i)$;
		\EndFor
	\end{algorithmic} 
\end{algorithm}

The iteration ($Irs$) is defined as:
\begin{equation}
Irs=\begin{cases}
			|h-h'|, & \text{if horizontal resizing}\\
            |w-w'|, & \text{if vertical resizing}
\end{cases}
\end{equation}, where $(h, w)$ is the original size, $(h', w')$ is the target size.\newline The energy map mentioned in \cite{avidan2007seam} is the so-called gradient map. It is typically calculated based on the equation:
\begin{equation}
E(I)=\left \vert\frac{\partial}{\partial x}I\right\vert +\left\vert \frac{\partial}{\partial y}I\right\vert
\end{equation}, where $I$ is the given input image. Although the gradient energy function has good efficiency for many images it may cause serious distortion due to the sensitivity to noise and allocation of higher energy to edge pixels. Therefore, this map itself is not sufficient to avoid significant distortion occurs in the seam carving method \cite{avidan2007seam}.

Contrast to the approach in \citet{avidan2007seam}, we adopt the proposed network to produce the energy map. The energy map at each iteration can be expressed as:
\begin{equation}
    E_i = g(I_i)
\end{equation}
, where $g(.)$ denotes the model of the proposed network, $I_i$ is the retargeted image at iteration $i^{th}$, $E_i$ is the energy map of $I_i$, $i\in[1\dots Irs]$. At each iteration, based on the generated energy map, seams are identified. A seam, which has the minimum energy, is defined and removed from $I_i$. At each iteration, once the energy map is produced through our model, seam carving \cite{avidan2007seam} is borrowed for seam identification and seam removal to produce final retargeting image.

\begin{figure}[hbt!]
  \centering
  \includegraphics[width=3.4in]{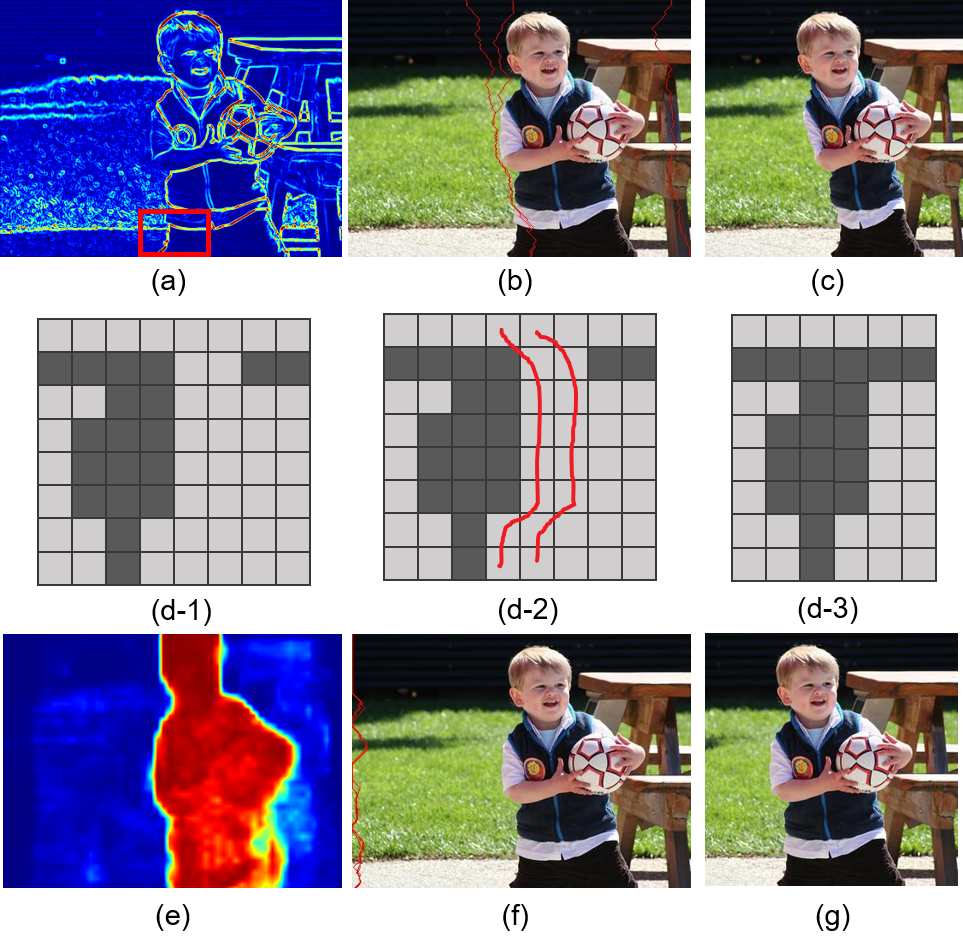}
  \caption{Illustrates the difference between our importance map and the gradient map in seam removal. (a) gradient map, (b) the three seams are removed, (c) retargeted result with the energy map (a). (d) visualization of pixel energy in the red rectangle in (a). The dark brown represents for high energy pixel. From left to right: energy map, the removed seams (red strokes), after removing seams. (e) our importance map, (f) the three seams are removed, (g) retargeted result with the energy map (e).}
  \label{f_seam_1}
\end{figure}

Fig.\ref{f_seam_1} illustrates the difference between our importance map and the gradient map in seam removal. Since the important regions of image are not presented with high energy in the gradient map, the removed seams cut through the objects (as in Fig.\ref{f_seam_1}b). Meanwhile, the importance map produced from our network controls efficiently to avoid distortion when removing seams. Therefore, the retargeted result is significantly improved in our result (g) compared to \citet{avidan2007seam}'s result (c).
\begin{figure}[hbt!]
  \centering
  \includegraphics[width= 3.4in]{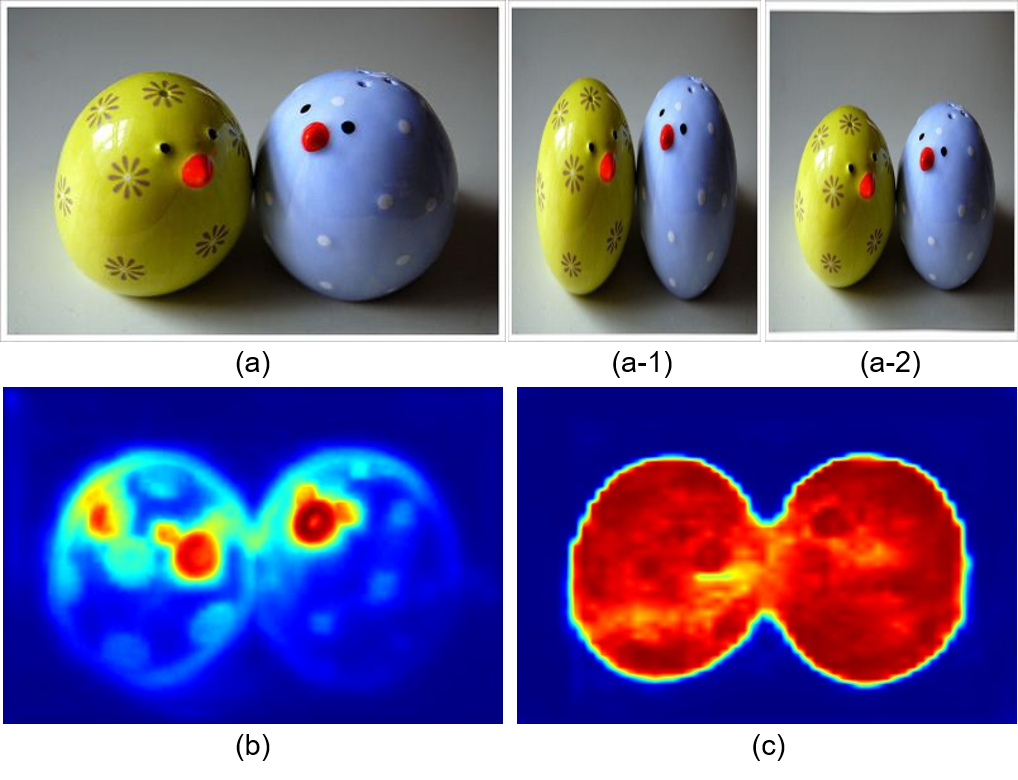}
  \caption{Efficiency of importance map generated from our proposed network in retargeting image. (a) Input image, (b) heat map of the saliency map obtained in \cite{goferman2011context}, (c) heat map of the importance map obtained in our network. (a-1) image is retargeted with the importance map (b), (a-2) image is retargeted with the importance map (c).}
  \label{f_result_map}
\end{figure}

\textbf{Image and video retargeting with patch-based approach}

The patch-based approach for image and video retargeting is introduced by \citet{lin2012patch} and \citet{lin2013content}, respectively. This technique firstly take advantage of a saliency detection method \cite{goferman2011context} to indicate the important region in image/video frames. For short, we use the terms \say{input} to imply input image and video frames in the following. The input is then segmented \cite{felzenszwalb2004efficient, grundmann2010efficient} into patches. The average of saliency value of each patch is computed. These values are then used in the optimization equations which are raised to preserve the shape of objects and the line structure after image/video is resized. This approach typically has a good performance in many images and videos which contain various attributes in the content. However, in the cases of the visual values of distinct regions that are similar or the content of the input is made up of sizable objects, the mentioned saliency detection method \cite{goferman2011context} fails in assisting the patch-based method \cite{lin2012patch, lin2013content} to generate the good retargeting results. In the following, we present how we tackle this limitation through our proposed network.

The given color input is segmented into $n$ patches $P = \{p_1, \dots , p_n\}$. Each patch is assigned an energy value $(\omega)$, which is defined as:
\begin{equation}
    \omega_k = \frac{1}{m}\sum_{j=0}^m s_{j}
\end{equation}, where $m$ denotes the total number of pixels in patch $p_k$, $s_{j}$ is the visual information value at a pixel in patch $p_k$; $k \in [1\dots n]$.

In \citet{lin2012patch, lin2013content}, the energy values of patches, which is defined in Eq(9), are used in warping. A high energy value is assigned to nonhomogeneous regions, whereas a low energy value is assigned to homogeneous regions. The saliency detection method \cite{goferman2011context} is used to detect the object regions, especially the boundary of regions. However, the inner region, which locates inside the important object, is not presented sufficiently to be recognized as an important region.

\begin{figure}[hbt!]
  \centering
  \includegraphics[width=3.0in]{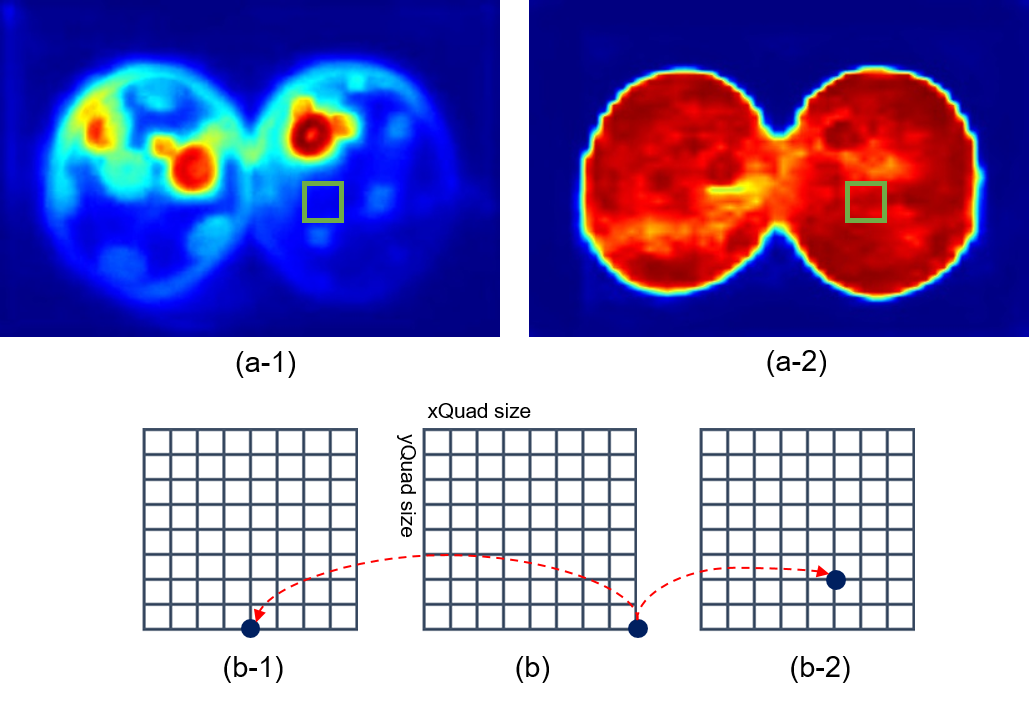}
  \caption{(a-1) the heat map of visual information produced by \cite{goferman2011context}; (a-2) the heat map of our importance map; (b), (b-1), (b-2) is a the visualization of a quad in the green region in (a-1) and (a-2). And, the blue circle represents for a vertex of the quad. (b) coordinate of quad vertex in original image. (b-1) predicted coordinate of quad vertex in retargeted result using (a-1). (b-2) predicted coordinate of quad vertex in retargeted result using (a-2).}
  \label{fig_7}
\end{figure}

To tackle this problem, we utilize our proposed network to generate an importance map which is then used to calculate the energy value on each patch through Eq(9) for both images and video frames. In Fig.\ref{fig_7}, the heat map of the saliency map \cite{goferman2011context} (shown in (a-1)) and our importance map (shown in (a-2)) of the input image in Fig.\ref{f_result_map}-a are presented. We can see in the same region, which is highlighted by the green square, it is recognized as unimportant region in a-1 but it has a high energy in (a-2). Thus, when \citet{lin2012patch, lin2013content} employ the saliency map \cite{goferman2011context} in their retargeting system, vertices in the quad, which belong to such regions, are scaled uniformly (as shown in (c-1)). This phenomenon results in the important objects are shrunk after resizing (as in Fig.\ref{f_result_map}a-1). With the utilization of our network, the visual information of considerable objects is sufficiently defined to describe the property of such regions. Thus, vertices in the quads in these regions are deformed ununiformly in the consistency with the significance of the corresponding region (as shown in (c-2)). This aspect boosts the shape of the important objects that can be retained close to the ratio of the original ones (as shown in Fig.\ref{f_result_map}a-2). 

Moreover, in terms of preserving the shapes of high-significance patches, \citet{lin2012patch, lin2013content} deal with this issue by defining the total patch transformation energy with a weighting factor assigned to each energy term. 
\begin{equation}
\begin{split}
    D_{TF}(P) &= \sum_{k=1}^{np}(\alpha \times D_{ST}(patch_k) \\
    &+ (1-\alpha) \times D_{LT}(patch_k))
\end{split}
\end{equation} 

\begin{equation}
\begin{split}
    D_{Sp}(M) &= (\alpha \times D_{SimT}(M) + (1-\alpha) \times D_{LinT}(M)) \\
    &+ D_{Ori}(M)
\end{split}
\end{equation}

The equation (10) and equation (11) is called the total patch transformation energy in \cite{lin2012patch} and \cite{lin2013content}, respectively. In these equations, a large value of the weighting factor $\alpha$ is assigned $(\alpha \in \{0.7, 0.8\})$. In this manner, the shape of the important objects can be preserved but they are still shrunk. Assuming $\omega_k'$ is the energy of patch $k$, which is produced by: $\omega_k' = \alpha \times \omega$. $\omega_k'$ grows linearly  with the energy of other patches in the same factor. Thus, there is a lack of correlation between patches. In our current system, since the importance map generated by our network is sufficient to describe the visual information of the input, we propose to modify the total patch transformation energy in \cite{lin2012patch, lin2013content}. That is, we eliminate the weighting factor $\alpha$ in our modified versions. We define as followed:
\begin{equation}
    E(P) = \sum_{k=1}^n(e_t(p_k) + e_s(p_k))
\end{equation}, where $e_t, e_s$ is the energy term of rigid transformation and linear scaling \cite{lin2012patch} of patch $p_k$, respectively; $k \in [1 \dots n]$, $n$ is the total number of patches in a segmented input. 
\begin{equation}
    E(M) = e_{lim}(p_k) + e_{lin}(p_k) + e_{ori}(p_k)
\end{equation}, where $e_{lim}, e_{lin}, e_{ori}$ is the energy term of rigid transformation, linear scaling, and grid orientation \cite{lin2013content}, respectively, in patch $p_k$; $k \in [1 \dots n]$; $n$ is the total number of patches in segmented frame. And, all of the energy terms in Eq(12) and Eq(13) are adopted from our importance map.
\begin{figure}[hbt!]
  \centering
  \includegraphics[width=3.4in]{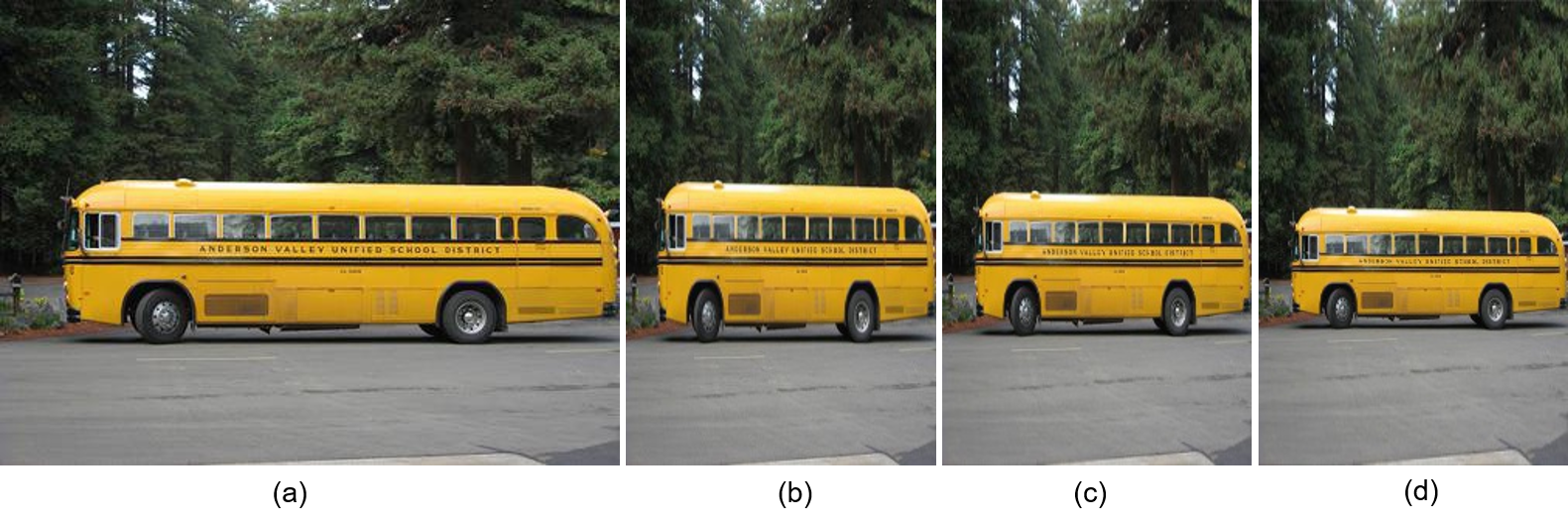}
  \caption{(a) Original image, \citet{lin2012patch}'s result without weighting factor $\alpha$ in Eq(10) (b) and with Eq(10) (c), our result with Eq(12) (d). In this example, the original image is reduced 50\% of original width.}
  \label{fig_factor_alpha}
\end{figure}

We describe the capability of Eq(12) and Eq(13) in preserving the shape of important objects in comparison with Eq(10) and Eq(11) in Fig.\ref{fig_factor_alpha}. We can see in this figure if the authors in \cite{lin2012patch,lin2013content} do not employ a weighting factor to increase the energy of significant patches the shape of the important object is stretched seriously (shown in (b)). When assigning the weighting factor to each energy term (as in Eq(10)), the shape of object is just slightly improved (shown in (c)). In contrast, with our importance map and Eq(12), our result (shown in (d)) has a better performance in terms of preserving the shape and the ratio of the important object.

\begin{figure}[hbt!]
  \centering
  \includegraphics[width=3.3in]{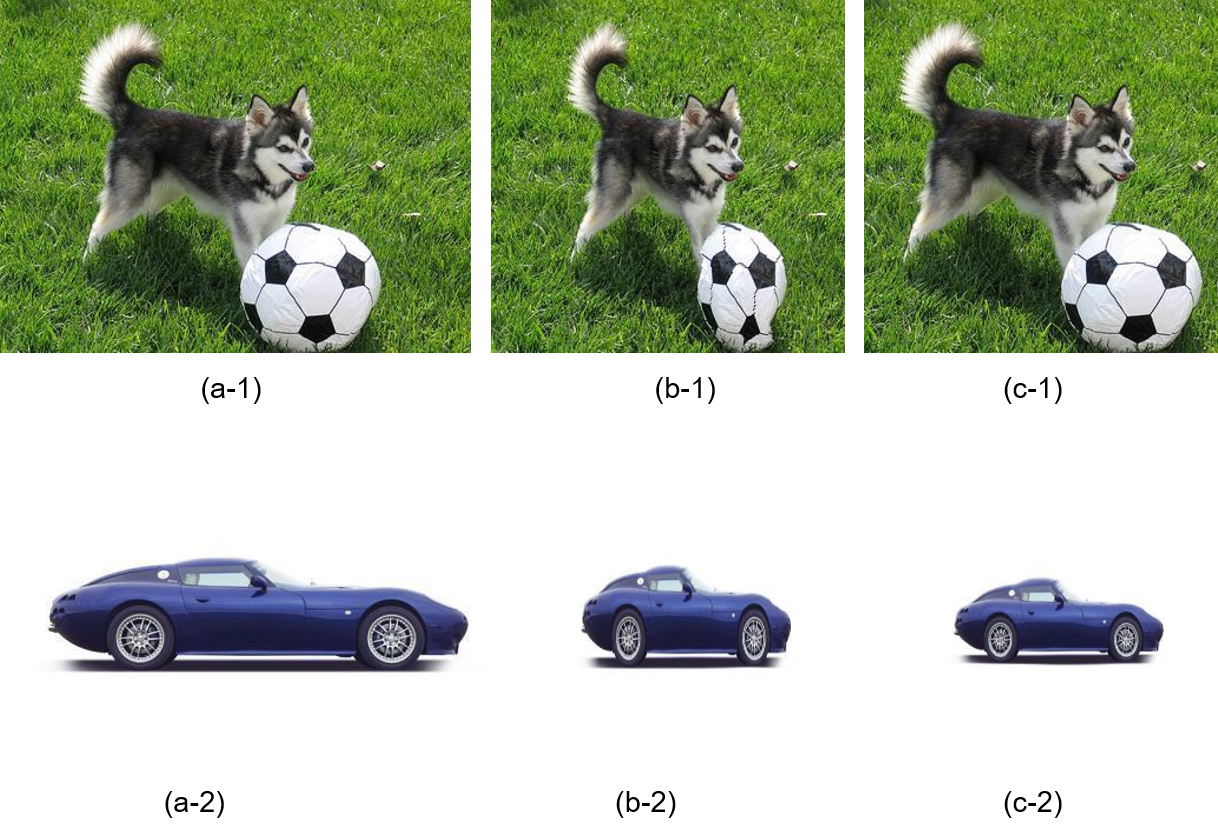}
  \caption{(a) original images; retargeting image (a-1) with seam carving operator by gradient map (b-1) and our importance map (c-1); retargeting image (a-2) with patch-based method by saliency map (b-2) and our importance map (c-2).}
  \label{fig_result_1}
\end{figure}

\begin{figure}[hbt!]
  \centering
  \includegraphics[width=3.0in]{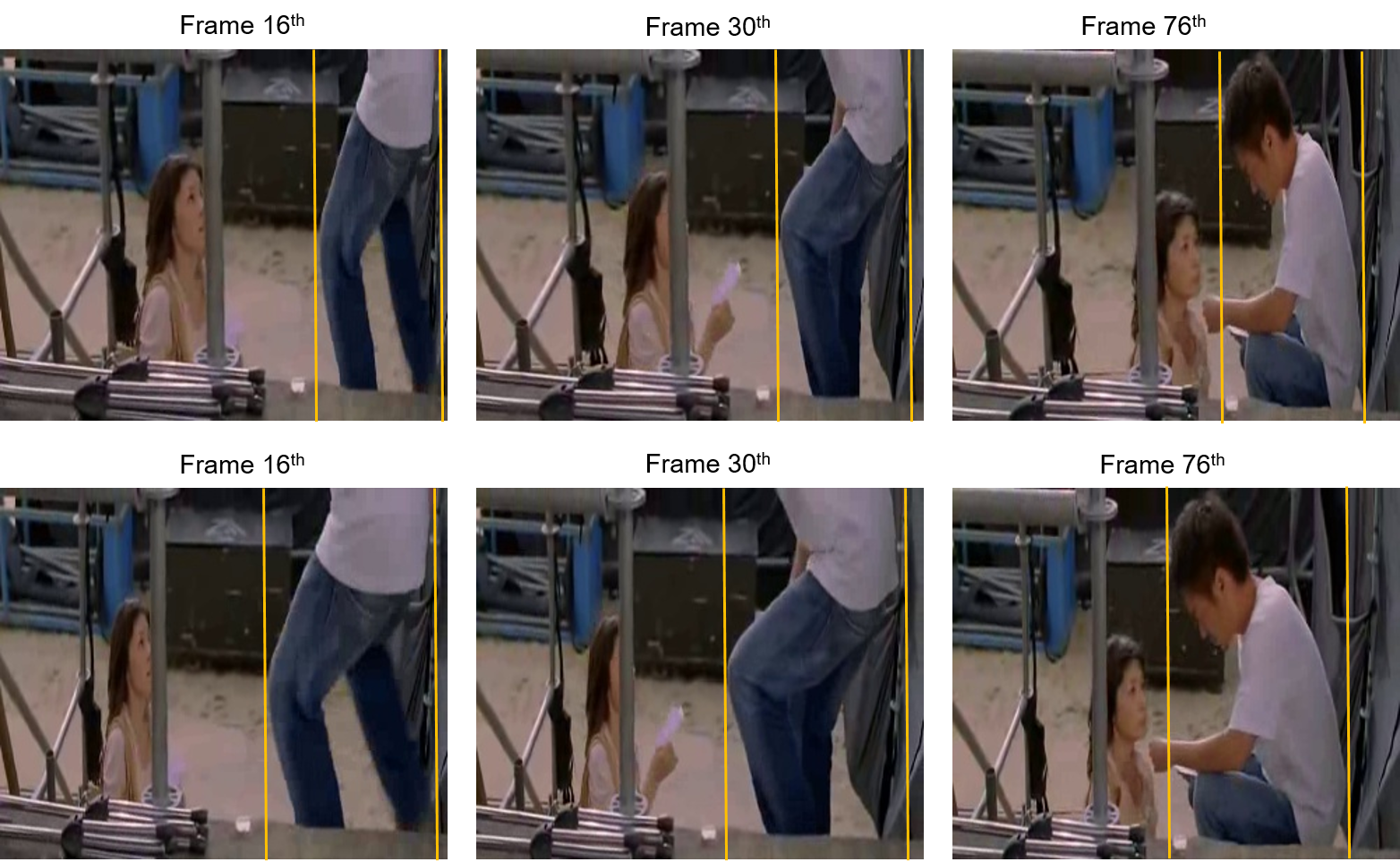}
  \caption{Each pair of frames, which are extract from different frames in the same video, is compared in each column. The first row is retargeted frames obtained from \citet{lin2013content}. The second row is the retargeted frames generated by our system.}
  \label{fig_result_2}
\end{figure}

After employing our importance map to calculate the energy of each patch in image/frames, the studies \cite{lin2012patch, lin2013content} are borrowed to produce retargeted results. We regard to note that the modified versions of patch transformation energy, which are presented in Eq(12) and Eq(13), are used in this process instead of Eq(10) and Eq(11). Figures \ref{fig_result_1} and \ref{fig_result_2} show the effectiveness of our media retargeting system in handling the limitation in both seam carving and patch-based techniques through our better results. The other results, we present in the next section.

\section{Experimental Results}
\begin{figure*}
  \centering
  \includegraphics[width=2\columnwidth]{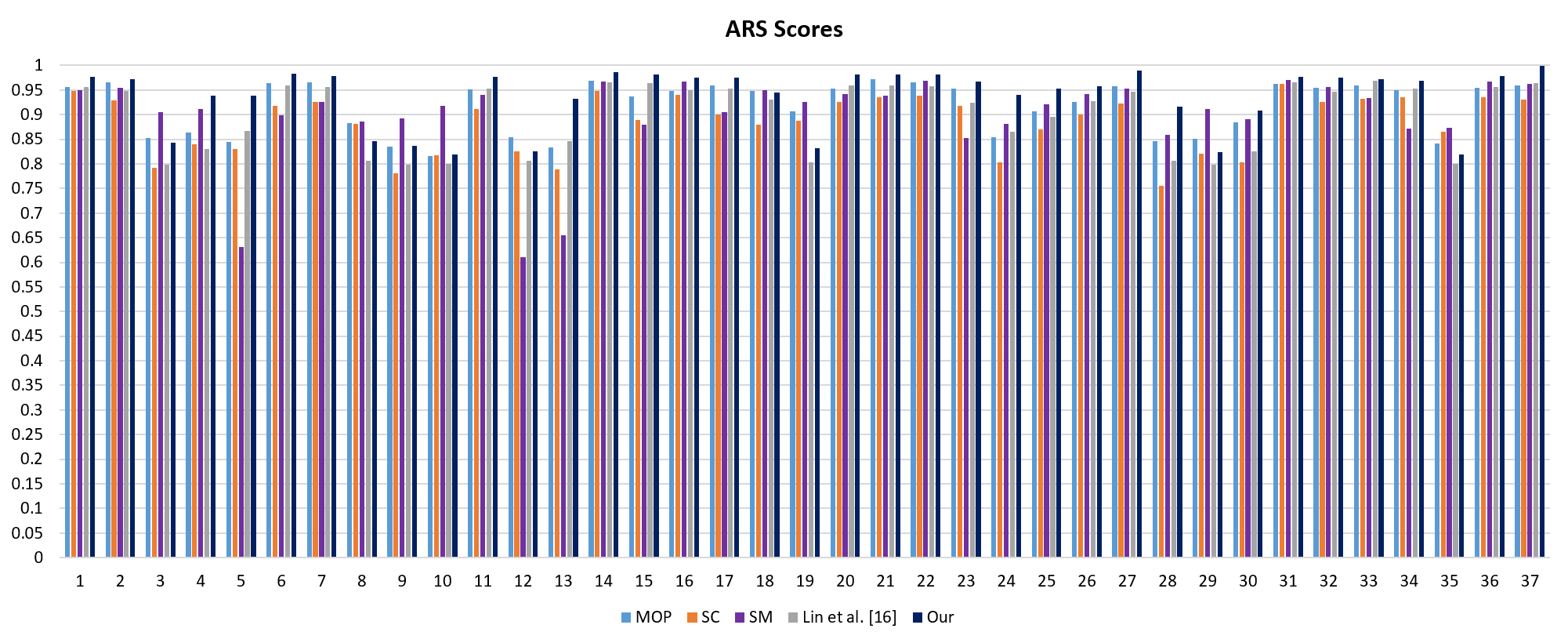}
  \caption{The evaluation on the ARS score.}
  \label{f_ars_score}
\end{figure*}
We implemented our system on the PC with Intel Core i7 CPU, 16GB RAM, and Nvidia GTX1070 GPU. With an input image/frame in the size of 1024x768, our proposed model takes less than one second to generate an importance map. Meanwhile, the algorithm in \cite{goferman2011context} takes about 45 seconds to produce a saliency map. In our overall system, we implement the proposed network with Python version 3.6, the segmentation process and the retargeting operators are computed with C++ programming language in Visual Studio 2015. In our proposed network, we use MSRA 10K dataset \cite{liu2010learning}, which is used for saliency detection, as our training data. This dataset consists of 10000 images with the diversity of the content structure of natural scenes. The dataset also contains manually annotated ground-truth saliency.

\subsection{Objective evaluation}
To quantitate the quality of our retargeting results, we adopt the Aspect Ratio Similarity (ARS) score. This measurement is introduced in \cite{zhang2016aspect} and has been mentioned as an effective method to estimate the geometric transition for the relationship images \cite{song2018carvingnet, kiess2018survey}. Thus, this assessment is suitable to adequately measure the aspect ratio changes between the original and retargeted images. 

In this computation, RetargetMe dataset \cite{rubinstein2010comparative} is used. There are a total of 37 sets of images with different attributes in image content. Each set consists of an original image and the corresponding retargeted images which are retargeted by eight retargeting methods \cite{rubinstein2010comparative}. In our objective evaluation, we conduct this measurement on our image retargeting results in comparison with four typical retargeting methods' results. They are Multi-operators (MOP) \cite{rubinstein2009multi}, Seam carving (SC) \cite{avidan2007seam}, Shift-map (SM) \cite{pritch2009shift}, and Warping \cite{lin2012patch}. All of the results in this dataset are published by the authors, thus they are trustful enough to use and for a fair comparison. The ARS score of each method is normalized in the range from 0 to 1. That is, a method that has a higher ARS score implies the retargeted result generated from that method is better.

\subsection{Discussion}
\begin{figure}[hbt!]
  \centering
  \includegraphics[width=3.3in]{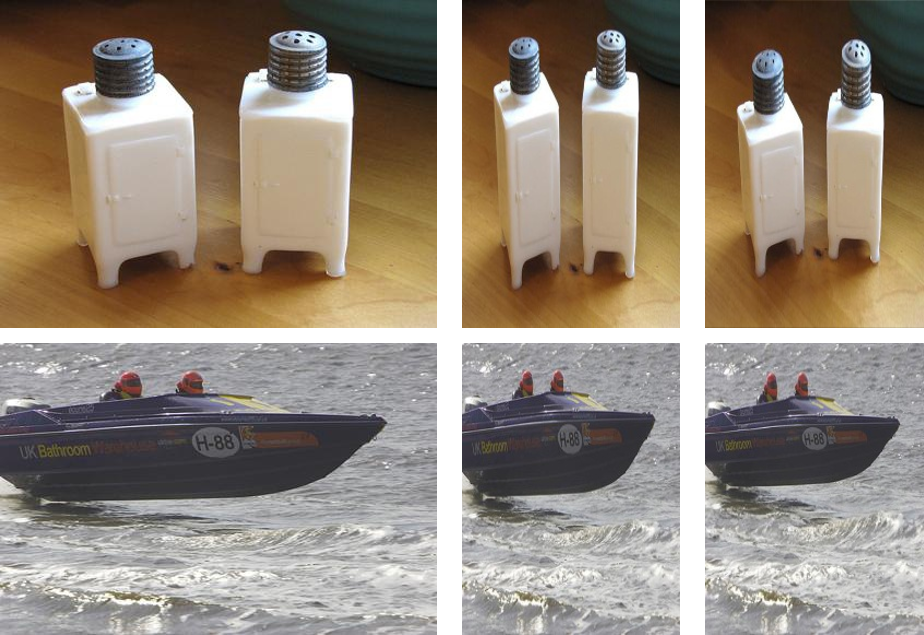}
  \caption{Images are retargeted by patch-based method. From left to right: original images, results in \cite{lin2012patch}, our results.}
  \label{fig_results_images}
\end{figure}

\begin{figure*}
  \centering
  \includegraphics[width=2\columnwidth]{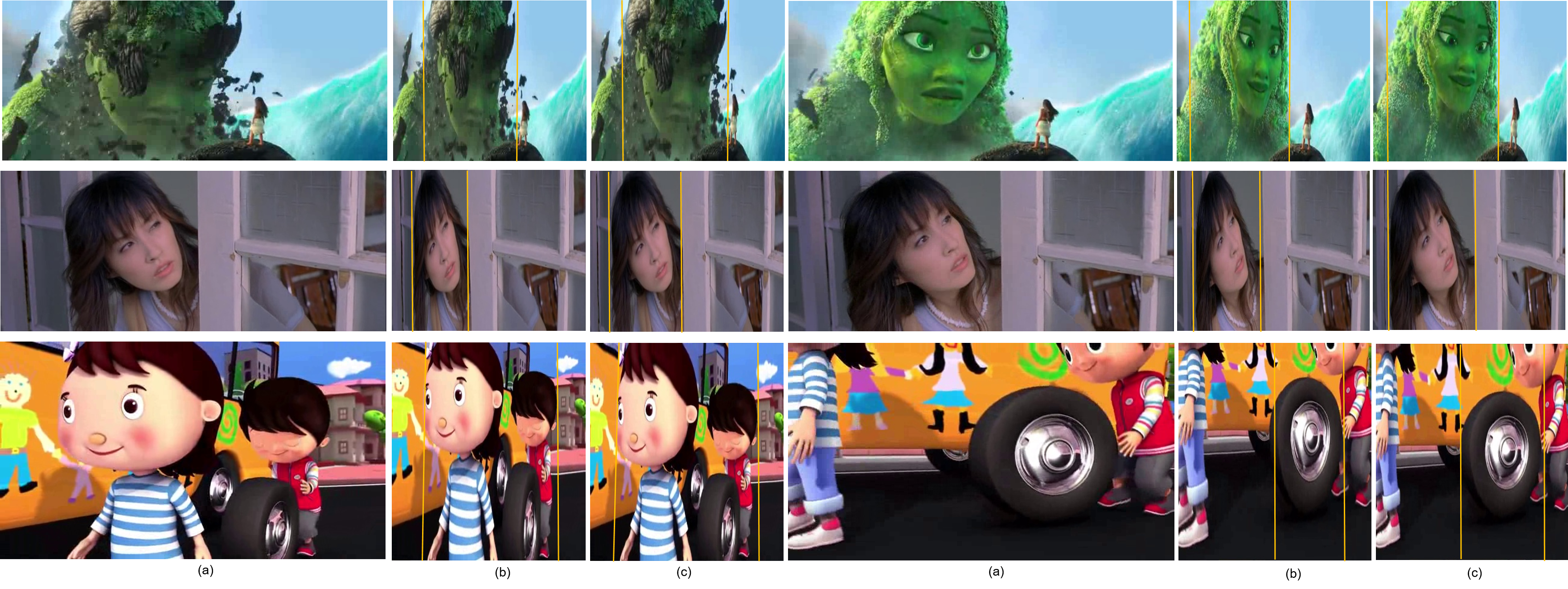}
  \caption{Examples of retargeted videos. (a) Original frame, (b) \citet{lin2013content}'s result, (c) our result.}
  \label{fig_results_videos}
\end{figure*}

In terms of quantitating our results in comparison with the typical retargeting methods on the benchmark dataset RetargetMe, we show the objective evaluation in Fig.\ref{f_ars_score}. Based on the ARS scores in this figure, most of our retargeting results have a good quality and better than the compared results. There are 8 cases (21\%), the scores are not the highest in the set, but they are still in an acceptable level (greater than 0.8) and higher than \citet{lin2012patch}'s results. 
\begin{figure}[hbt!]
  \centering
  \includegraphics[width=3.3in]{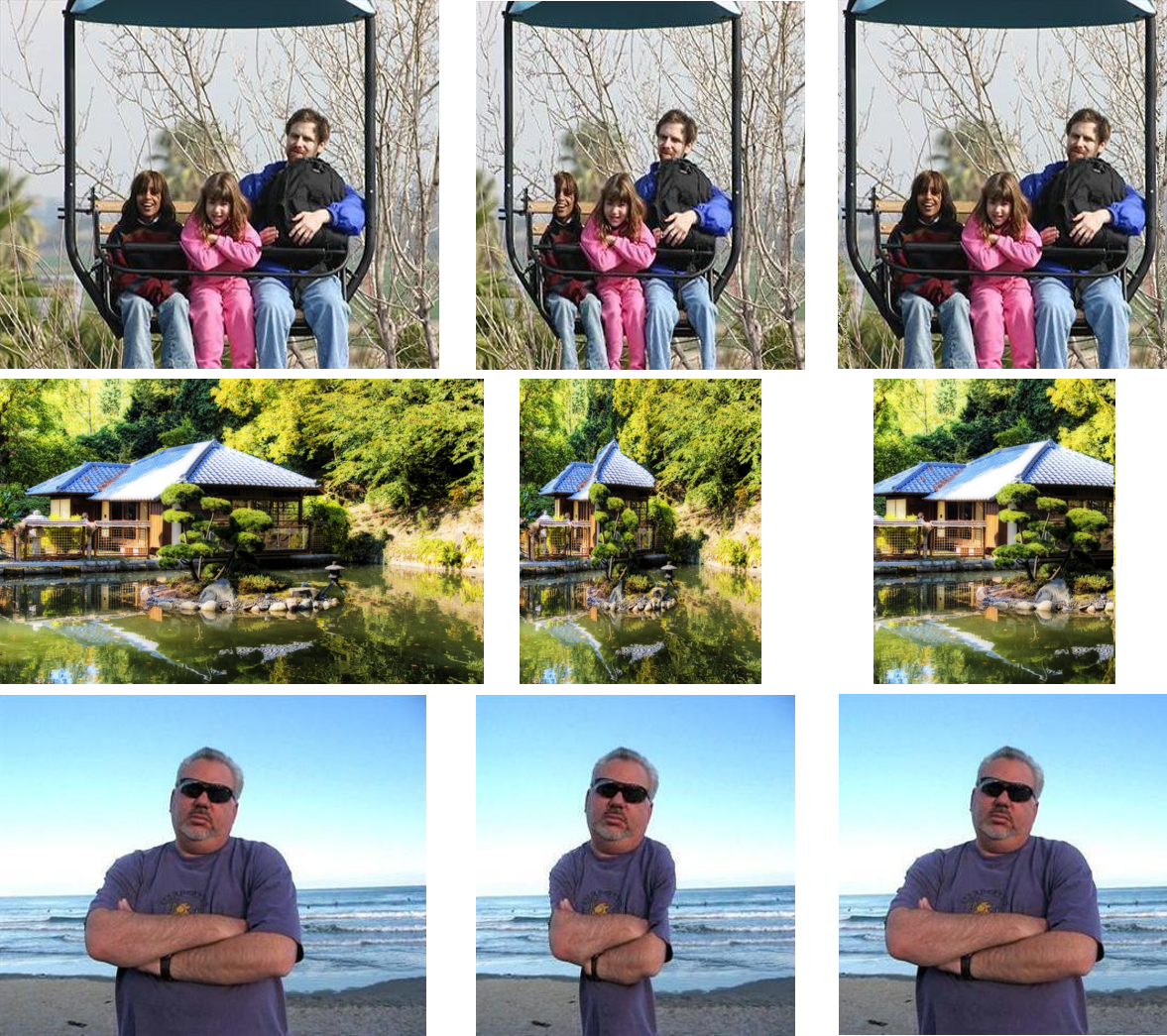}
  \caption{Images are retargeted by seam carving operator. From left to right: original images, results in \cite{avidan2007seam}, our results.}
  \label{fig_results_sc}
\end{figure}
Besides the RetargetMe dataset, we conduct more experiments on other data (NRID \cite{hsu2014objective} and DUTS-TR dataset\footnote{http://saliencydetection.net/duts}). We show respectively our retargeted images and videos in  comparisons with the studies \citet{avidan2007seam} in Fig.\ref{fig_results_sc}, \citet{lin2012patch} in Fig.\ref{fig_results_images}, and \citet{lin2013content} in Fig.\ref{fig_results_videos}. In term of seam carving operator, we examine on the original images which contain different attributes in their content, as shown in Fig.\ref{fig_results_sc}. The results in \cite{avidan2007seam} are distorted seriously with the utilization of the gradient map. Whereas, our results preserve effectively most of the important regions in the original images. For the warping based method, Figures \ref{fig_results_images} and \ref{fig_results_videos} present our results in the comparisons with \cite{lin2012patch} and \cite {lin2013content}, respectively. As shown in these two figures, the retargeted images and frames produced in \cite{lin2012patch, lin2013content} are shrunk. However, our importance map boosts these results quite better. 

\begin{figure}[hbt!]
  \centering
  \includegraphics[width=3.3in]{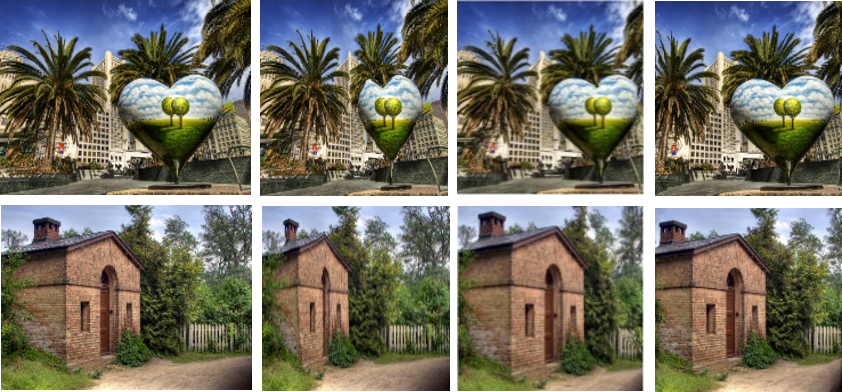}
  \caption{From left to right: original image, \citet{avidan2007seam}'s results, \citet{song2018carvingnet}'s results, our results with seam carving operator.}
  \label{fig_compare_1}
\end{figure}
\begin{figure}[hbt!]
  \centering
  \includegraphics[width=3.3in]{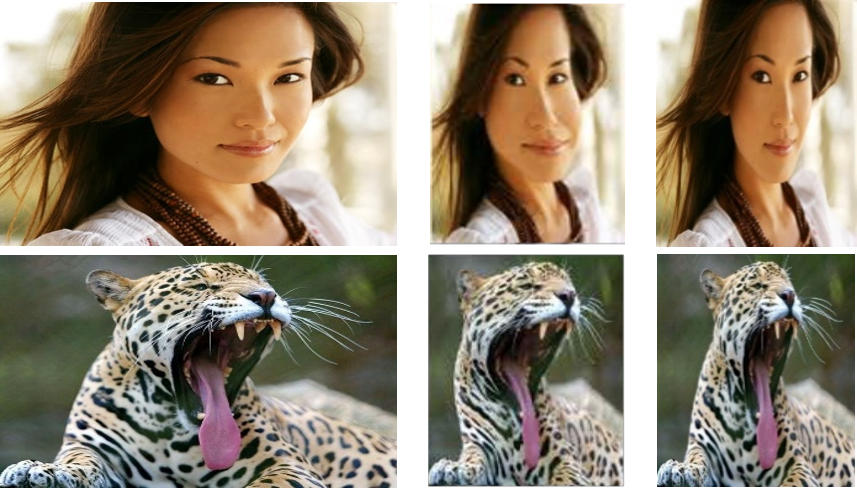}
  \caption{From left to right: original images, \citet{tan2019cycle}'s results, our results.}
  \label{fig_compare_2}
\end{figure}
\begin{figure}[hbt!]
  \centering
  \includegraphics[width=3.3in]{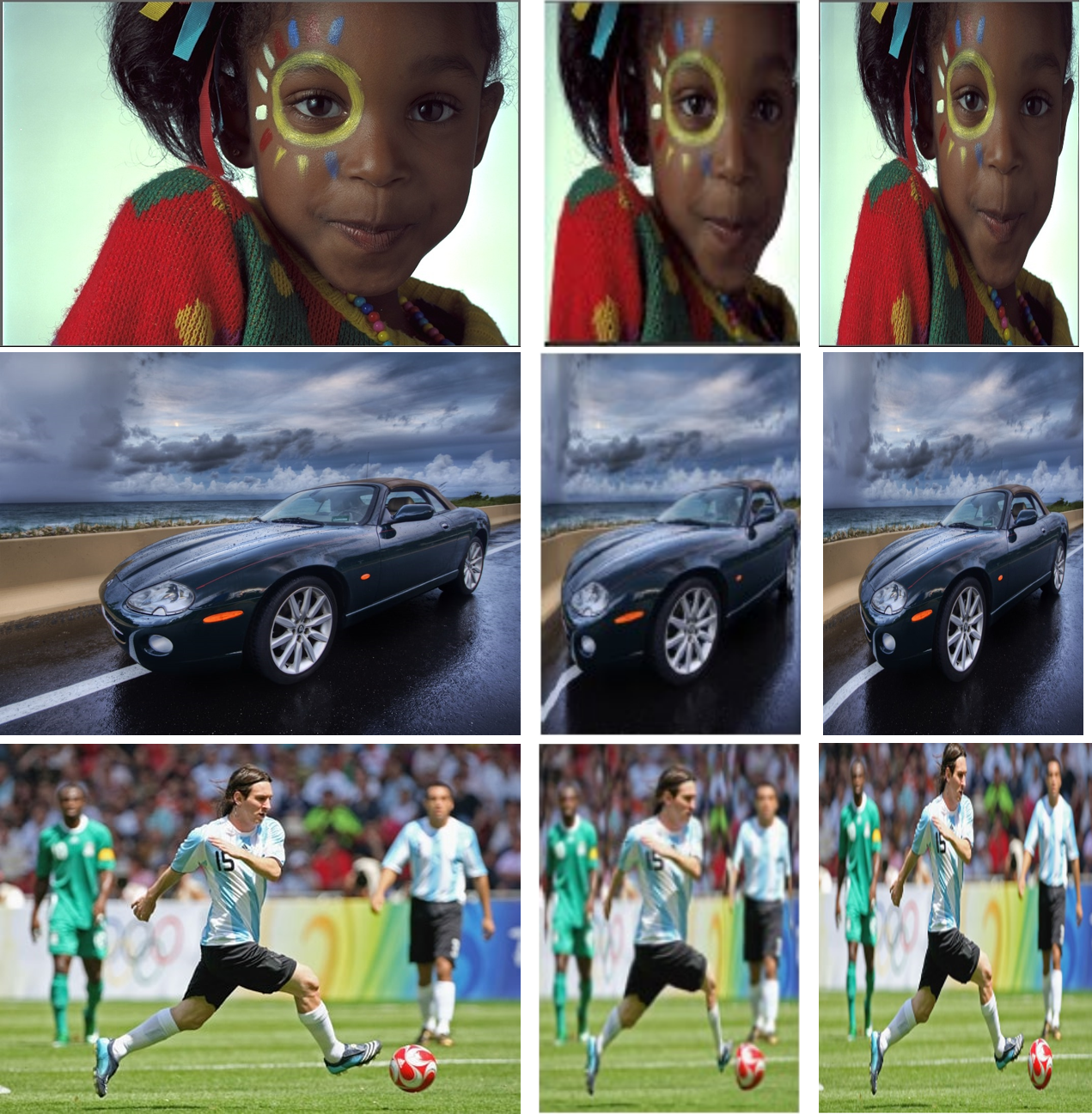}
  \caption{From left to right: original image, \citet{ahmadi2019context}'s result, our result.}
  \label{fig_compare_extra}
\end{figure}
\begin{figure}[hbt!]
  \centering
  \includegraphics[width=3.3in]{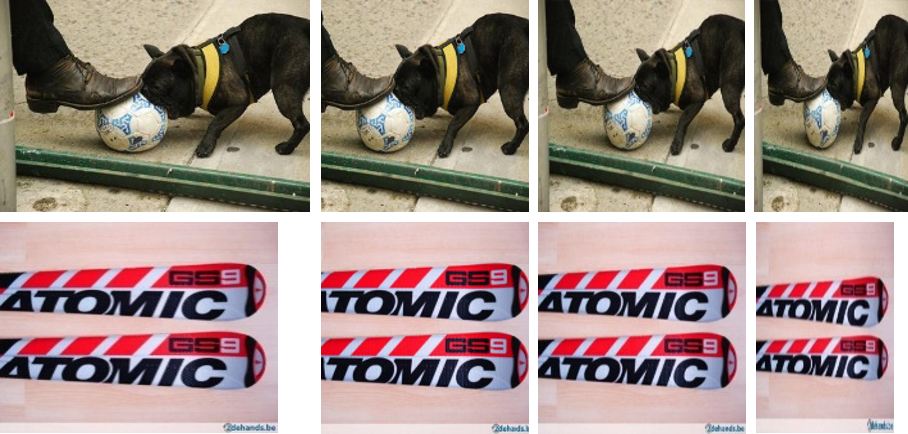}
  \caption{From left to right: original images, \citet{song2018photo}'s results with 25\% width reduction, our results with 25\% width reduction, our results with 50\% width reduction.}
  \label{fig_compare_3}
\end{figure}

Apart from the above, we also compare our results with those that are generated using deep learning. Fig.\ref{fig_compare_1} demonstrates the comparison of our results on seam carving operator with \citet{song2018carvingnet}. As shown in this figure, with the same operator (seam carving), the serious distortion occurs with the utilization of the gradient map in \cite{avidan2007seam} and deep energy map in \cite{song2018carvingnet}. In contrast, our results better preserve the content of the important regions including texture, edge, line, etc. Fig.\ref{fig_compare_2} shows our results in the comparison with \citet{tan2019cycle}. The retargeted images, which are generated through the network in \cite{tan2019cycle}, are significantly distorted. Meanwhile, ours have fully better performance. Fig.\ref{fig_compare_extra} is presented to compare our results with retargeted images in \cite{ahmadi2019context}. The authors in \cite{ahmadi2019context} uses convolutional neural networks to generate saliency map for image retargeting. From these results, their retargeted images are distorted but ours are successful. The reason they fail in these examples is that, as stated in \cite{ahmadi2019context}, the method they use assigns equal scaling factors to all the pixels in columns. Thus, by sudden changes in the scaling factor from a column to the next column, the straight line is deflected (as in the the car image). And, this reason also causes distortion in the less salient columns (as in the right ear of the child and the player's feet). Another deep learning-based study we compare in this sector is the method proposed in \cite{song2018photo}. This research proposes a network to drive an image to a square size. This comparison is depicted in Fig.\ref{fig_compare_3}. The results in \cite{song2018photo} are not distorted but they are cropped on the left side of images. This phenomenon leads to the squared photos fail in preserving image content in the cases that the important content of the image are not located in the center area. However, we handle well in these examples even they are retargeted to 50\% of the width. 
 

\begin{figure}[bht!]
  \centering
  \includegraphics[width=3.3in]{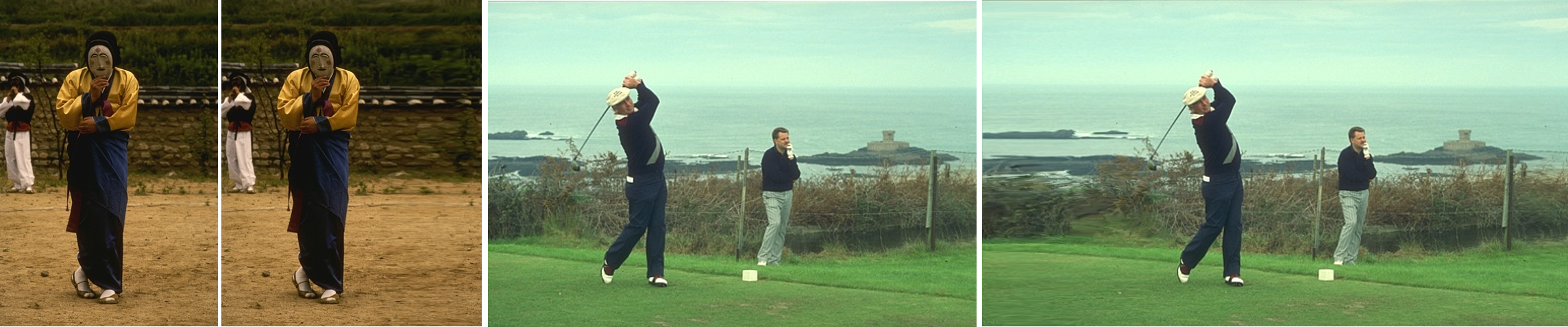}
  \caption{Enlarging images to 25\% of the width by our system. In each pair of images, the right image is enlarged from the left one.}
  \label{fig_enlarge}
\end{figure}

\begin{figure*}
  \centering
  \includegraphics[width=2\columnwidth]{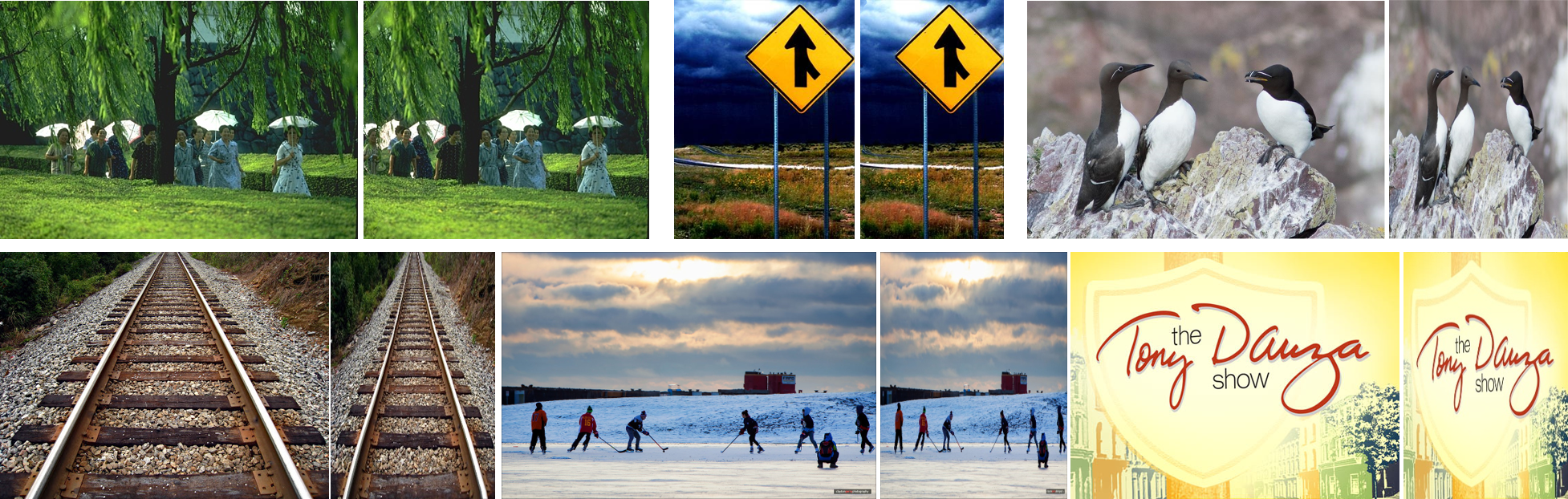}
  \caption{More results on different image attributes in our system. In each pair of images, the left image is original image of the retargeted result on the left.}
  \label{f_attributes}
\end{figure*}

To demonstrate more about the capability of our importance map in the retargeting applications, we examine to enlarge images by our system (shown in Fig.\ref{fig_enlarge}). Moreover, \citet{tang2019image} showed that the images with low retargetability scores are reliable for assessing the new retargeting method. Such images contain attributes like text, symmetry, texture, multiple objects. Motivated by this, we conduct the experiment on the images which are made of these attributes. Fig.\ref{f_attributes} shows the plausible results are generated by our system in this manner. This implies that we not only boost the retargeting result better but also handle well on the \say{difficult-to-resize} images. More experimental results including videos are supplied in our supplementary document which can be explored at our website:
\href{http://graphics.csie.ncku.edu.tw/deep_saliency}{http://graphics.csie.ncku.edu.tw/deep\_saliency}.

\begin{figure}[bht!]
  \centering
  \includegraphics[width=3.3in]{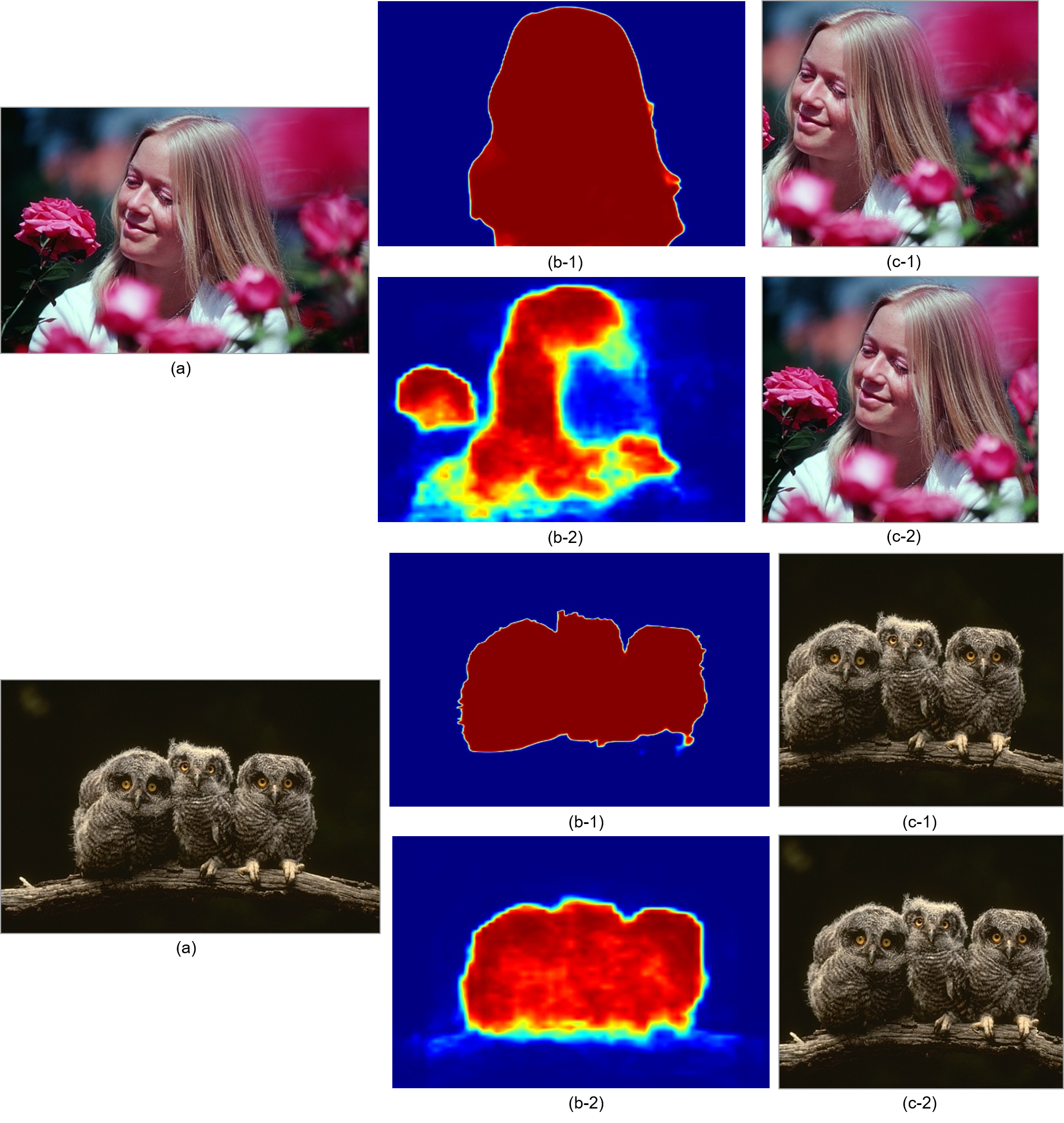}
  \caption{(a) original image, (b-1) saliency map in \cite{qin2019basnet}, (c-1) retargeted image with seam carving operator when using (b-1) as the energy map, (b-2) our importance map, (c-2) retargeted image with seam carving operator when using (b-2) as the energy map.}
  \label{fig_bas}
\end{figure}

In the aforementioned part, we mostly compare our work with the related works of different classes in retargeting. The question here is what if the proposed importance map is substituted by another deep learning-based saliency map. We argue this issue through the results in Fig.\ref{fig_bas}. And, we examine the saliency map obtained from \cite{qin2019basnet} in this experiment. The BASNet, which is proposed in \cite{qin2019basnet}, has a good performance in salient object detection and has been used in many image processing applications. However, when it is employed in retargeting application, it may not be adequate (as the results in images c-1). The reason is that an importance map for retargeting application is required to represent the energy for all the pixels in an image. The magnitude of pixel energy describes how significant a pixel is. A good salient object detection method focuses on efficiently detecting the recognized objects (see images b-1). Thus, the pixels in other regions are skipped. This phenomenon results in the retargeted images distorted with seam carving operator or inconsistent deformation with warping method. 

\begin{figure}[bht!]
  \centering
  \includegraphics[width=3.3in]{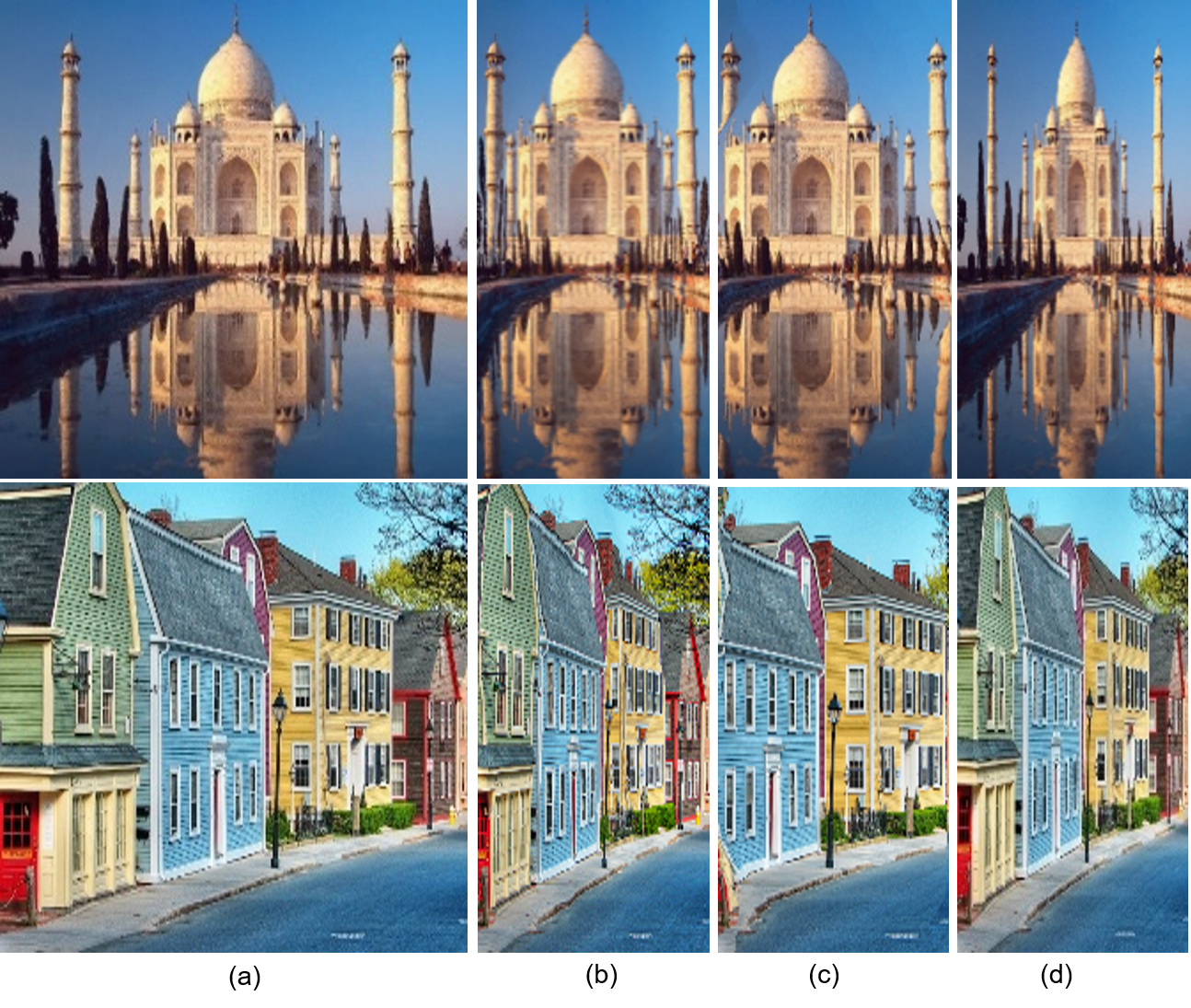}
  \caption{Our examples of failure with seam carving operator. (a) Original image, (b) \citet{avidan2007seam}'s result, (c) Our result with seam carving operator, (d) our result with warping method. The examples in this figure are retargeted to 50\% of the width.}
  \label{fig_fail_cases}
\end{figure}

We finally show our examples of failure in Fig.\ref{fig_fail_cases}. Due to the limitation in the dataset we use to train, the importance map obtained from our network is not good enough to represent the information on the input. Consequently, it may not perform well with seam carving operator. As shown in Fig.\ref{fig_fail_cases}, the results are not good with the utilization of seam carving method. Yet, they are still plausible in our system with warping operator \cite{lin2012patch}. This problem may be also due to the original limitation of seam carving method. Our proposed network can reduce the possibility of failure cases but can not solve its original limitation, i.e., seam carving may include partials of important areas in the carved seams.

\section{Conclusion}
In this paper, we have presented a network to produce an importance map, which is used in the image/video retargeting system to enhance the quality of the image/video after retargeted. Once the importance map is obtained from the proposed network, a retargeting operator can be adopted to generate retargeted image/video. Two methods are examined in this study including seam carving, modified versions of warping for image and video retargeting. We demonstrate the effectiveness of our framework by an objective evaluation and several ideal retargeted results. Experimental results prove that the proposed scheme performs well in comparisons with those of other works. In future work, we hope to improve the dataset to alleviate the limitation in this study. Besides, we plan to investigate new image retargeting and video retargeting methods using a deep learning-based approach.

\normalsize
\bibliography{references}

\end{document}